\documentclass[journal,12pt,onecolumn,draftclsnofoot]{IEEEtran}

\usepackage{color}
\usepackage{multirow}
\usepackage{graphicx}
\usepackage{epstopdf}
\usepackage[cmex10]{amsmath}
\usepackage{bm} 
\usepackage{amssymb}

\usepackage{multirow}
\usepackage{amsthm}
\usepackage{cite}
\usepackage{balance}
\usepackage{acronym} %This package will write out the acronym the first time it is used and write it as an acronym for the rest of the paper
\usepackage{algorithm, algcompatible}

\newcommand{\diag}{\mathop{\rm diag}}

\algnewcommand\INPUT{\item[\textbf{Input:}]}
\algnewcommand\OUTPUT{\item[\textbf{Output:}]}

\allowdisplaybreaks

\begin{document}

\title{Impact of Inter-Operator Interference via Reconfigurable Intelligent Surfaces} 
\author{Nikolaos~I.~Miridakis,~\IEEEmembership{Senior Member,~IEEE}, Theodoros~A.~Tsiftsis,~\IEEEmembership{Senior Member,~IEEE}, Panagiotis~A.~Karkazis, Helen~C.~Leligou, and Petar~Popovski,~\IEEEmembership{Fellow,~IEEE}
%\thanks{\textit{Corresponding Author: T. A. Tsiftsis.}}
\thanks{N.~I.~Miridakis and P.~A.~Karkazis are with the Department of Informatics and Computer Engineering, University of West Attica, Aegaleo 12243, Greece (e-mails: nikozm@uniwa.gr, p.karkazis@uniwa.gr).}
\thanks{T.~A.~Tsiftsis is with the Department of Informatics and Telecommunications, University of Thessaly, Lamia 35100, Greece (e-mail: tsiftsis@uth.gr).}
\thanks{H.~C.~Leligou is with the Department of Industrial Design and Production Engineering, University of West Attica, Aegaleo 12241, Greece (email: e.leligkou@uniwa.gr).}
\thanks{P. Popovski is with the Department of Electronic Systems, Aalborg University, Aalborg 9220, Denmark (email: petarp@es.aau.dk).}
}

%\markboth{}{}

\maketitle

\begin{abstract}
A wireless communication system is studied that operates in the presence of multiple reconfigurable intelligent surfaces (RISs). In particular, a multi-operator environment is considered where each operator utilizes an RIS to enhance its communication quality. Although out-of-band interference does not exist (since each operator uses isolated spectrum resources), RISs controlled by different operators do affect the system performance of one another due to the inherently rapid phase shift adjustments that occur on an independent basis. The system performance of such a communication scenario is analytically studied for the practical case where discrete-only phase shifts occur at RIS. The proposed framework is quite general since it is valid under arbitrary channel fading conditions as well as the presence (or not) of the transceiver's direct link. Finally, the derived analytical results are verified via numerical and simulation trial as well as some novel and useful engineering outcomes are manifested.  
\end{abstract}


\IEEEpeerreviewmaketitle

\section{Introduction}
\IEEEPARstart{R}{econfigurable} intelligent surfaces (RISs) have been extensively promoted lately to enhance the performance of beyond $5$G and $6$G communications (see, e.g., \cite{j:RenzoZappone20,j:alexandropoulos2023ris} and relevant references therein). Nonetheless, and rather surprisingly, the vast majority of research works so far focus on the analysis, modeling, and/or optimization of RIS-enabled communications based on the assumption that a single-only network operator controls and/or utilizes RIS(s). Such a system model assumption seems superficial when considering a multi-operator / multi-RIS environment. In practice, different wireless network operators coexist within a given geographical area, each functioning using distinct and thus non-overlapping frequency bands. Consequently, multiple users are simultaneously served by different operators in the (potentially close) vicinity. The case when a RIS deployment optimized by one operator to address the needs of its subscribed users causes an unavoidable impact on the performance of users served by other operators remains ambiguous. This ambiguity is particularly noteworthy as RIS elements are passive entities that reflect impinging signals across the entire spectrum of encountered frequencies.  

To date, quite a few research studies analyze the system performance under a multi-operator / multi-RIS environment \cite{j:CaiRang2022,j:Gurgunoglu2023,c:YashvanthMurthy2023,j:yashvanth2023}. In \cite{j:CaiRang2022}, the RIS phase shift configuration is jointly optimized in the presence of multiple operators; however, requiring inter-operator coordination which is not always feasible or desirable. In \cite{j:Gurgunoglu2023}, the important problem of pilot contamination during channel estimation in a multi-operator / multi-RIS environment is studied; yet, assuming a continuous phase shift design at RISs which is ideal. Most recently, \cite{c:YashvanthMurthy2023} and \cite{j:yashvanth2023} studied the performance of a multi-operator communication system; yet, assuming that only a single operator is equipped with a corresponding RIS technology. All these studies assumed either a multi-operator / single-RIS infrastructure and/or the absence of direct link channel gain and/or Rayleigh-only channel fading. To our knowledge, the performance analysis of a multi-operator / multi-RIS communication system under the presence of direct link and/or more general channel fading conditions is not available in the open technical literature so far.  

Capitalizing on the aforementioned observations, we analytically study the impact of inter-operator interference in the realistic condition of a multi-operator multi-RIS networking environment. As key performance indicators, system outage probability and spectral efficiency are obtained in straightforward closed-form expressions by assuming relatively large RIS arrays; which is practically feasible due to the passiveness of low-cost RIS elements. The derived results are valid for an arbitrary range of channel fading models (including the most popular ones such as Rician and Rayleigh fading) as long as they satisfy some mild conditions (defined in the next section). Also, the proposed model includes the presence (or not) of the direct link as well as the practical case of discrete-only phase shifts at the passive RIS elements \cite{j:MiridakisTsifYao2023}. Some key engineering insights are revealed, such as the system diversity order, influence of uncontrolled (multiple) RIS(s) on the total communication quality, while new closed-form expressions (generic and simplified asymptotic ones) of the said performance indicators are derived.

\section{System and Signal Model}
\begin{figure}[!t]
\centering
\includegraphics[trim=.5cm .5cm .5cm 0.0cm, clip=true,totalheight=0.4\textheight]{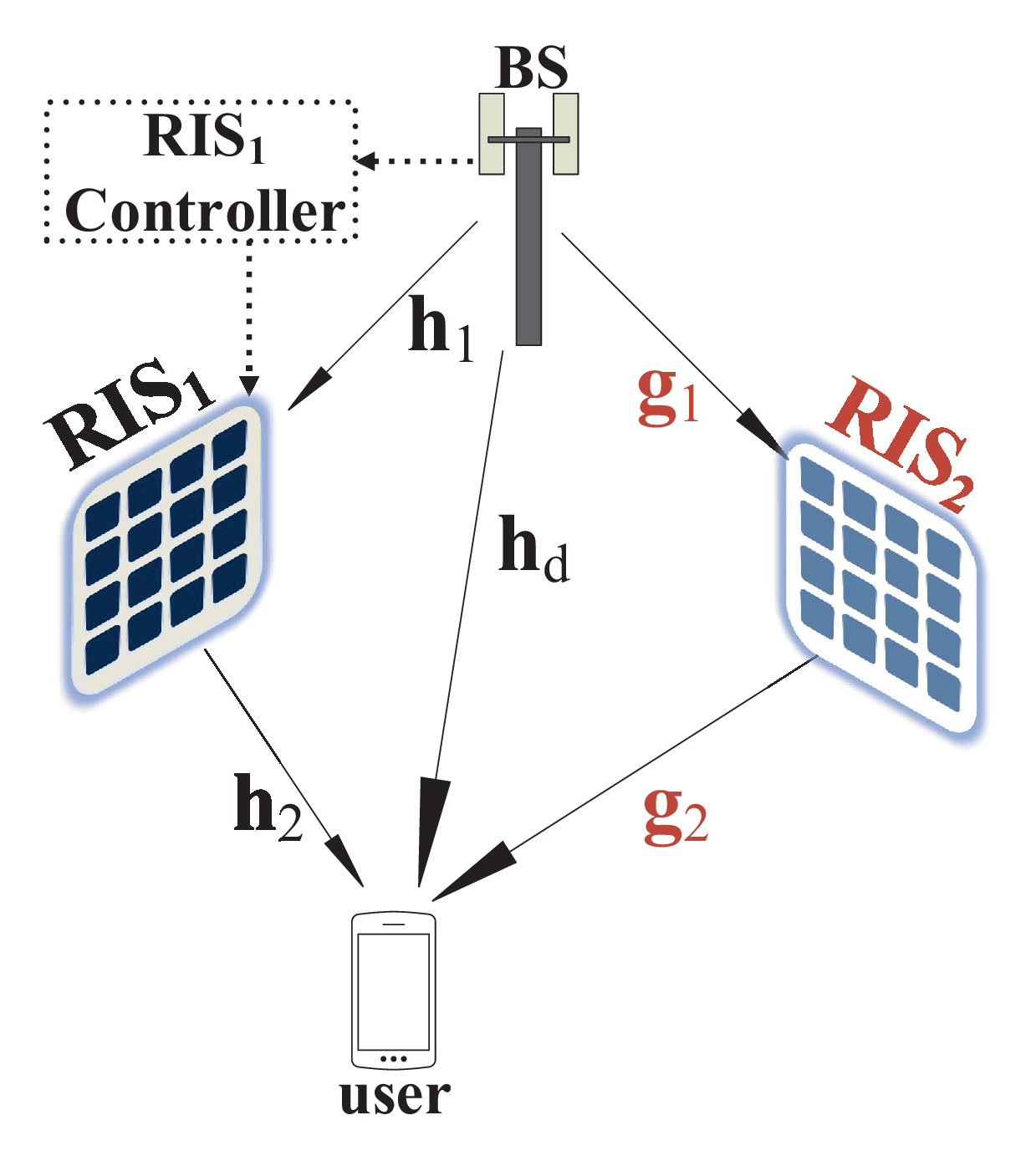}
\caption{The considered system model, where a base station (BS) stands for the transmitter, the receiver is the reference system user and the communication is aided through ${\rm RIS}_{1}$ which is directly controlled by its corresponding operator. Meanwhile, ${\rm RIS}_{2}$ operates \emph{independently} in a close vicinity, which is controlled by a different operator.}
\label{fig1}
\end{figure}

Consider a wireless communication system with a single-antenna transmitter and receiver operating over a quasi-static block-fading channel. The end-to-end communication is assisted by an intermediate RIS equipped with $N$ passive elements. It is also assumed that both the size of each element and the inter-element spacing are equal to half of the signal wavelength; such that the associated channels undergo independent fading \cite{j:RenzoZappone20}.\footnote{According to \cite{j:EmilLuca2021}, channel fading (spatial) correlation is always present in practical RIS illustrations. However, it was recently proved in \cite[Prop.~6 and Fig.~7]{j:WangBadiu2022} that, for a large-scale RIS (which is feasibly the practical case), the effect of spatial correlation introduces negligible impact to the system performance compared to the spatially independent assumption. Thus, to facilitate the following analysis, independent channel-faded links are considered hereinafter.} It is assumed that the transmitter-to-RIS and RIS-to-receiver links undergo independent Rician channel fading due to their relatively close distance and the (potential) presence of a strong line-of-sight (LoS) channel gain component.\footnote{The analysis can be extended to an \emph{arbitrary} channel fading type of the transmitter-RIS-receiver link as long as its underlying distribution is continuous and its first two moments are finite.} On the other hand, the transceiver link is subject to independent Rayleigh channel fading, due to a relatively high link distance and the presence of intense signal attenuation in a rich scattering environment. In addition, another operator coexists in a close vicinity of the considered system model, which utilizes a RIS-enabled communication for its own subscribers; viz. Fig.~\ref{fig1}. Although the two operators in principle occupy different spectrum bands and there is no out-of-band interference, the aforementioned infrastructure includes RISs with different features whereby a new type of interference inherently emerges; entitled as \emph{inter-operator interference} (IOI). It is caused by rapidly sudden changes in the phase shifts of the uncontrolled RIS (say, ${\rm RIS}_{2}$ of Fig.~\ref{fig1}), which unavoidably result to an incoherent sum of the received channel gain at the reference user. Due to the possibly different array sizes and mutually independent (uncontrolled) phase shifts of ${\rm RIS}_{1}$ and ${\rm RIS}_{2}$, we hereafter term these RISs as \emph{heterogeneous}.   

More specifically, the received signal reads as\footnote{{\textbf{Notation}:} Vectors and matrices are represented by lowercase and uppercase bold typeface letters, respectively. A diagonal matrix with entries $x_{1},\cdots,x_{n}$ is defined as $\diag\{x_{i}\}^{n}_{i=1}$. Superscript $(\cdot)^{T}$ denotes transpose; $|\cdot|$ represents absolute value, ${\rm Re}\{\cdot\}$ and ${\rm Im}\{\cdot\}$ is the real and imaginary part of a complex value, respectively, $\angle[\cdot]$ is the phase of a complex argument, and ${\rm j}\triangleq \sqrt{-1}$. $\mathbb{E}[\cdot]$ is the expectation operator, symbol $\overset{\text{d}}=$ means equality in distribution and $\overset{\text{d}}\approx$ defines almost sure convergence (asymptotically) in distribution. $f_{X}(\cdot)$ and $F_{X}(\cdot)$ represent the probability density function and cumulative distribution function (CDF) of a random variable (RV) $X$, respectively. $\mathcal{CN}(\mu,v,\rho)$ defines a complex-valued Gaussian RV with mean $\mu$, variance $v$ and pseudo-variance $\rho$, while $\mathcal{CN}(\mu,v)$ is the classical complex-valued Gaussian RV (with $\rho=0$). Also, $\Gamma(\cdot)$ denotes the Gamma function \cite[Eq. (8.310.1)]{tables}, $\Gamma(\cdot,\cdot)$ denotes the upper incomplete Gamma function \cite[Eq. (8.350.2)]{tables}, $\psi(\cdot)$ is the Digamma function \cite[Eq. (8.360.1)]{tables}, ${}_1F_{1}(\cdot,\cdot;\cdot)$ is the Kummer's confluent hypergeometric function \cite[Eq. (9.210.1)]{tables}, ${\rm sinc}(x)=\sin(x)/x$ is the Sinc function, $Q_{1}(\cdot,\cdot)$ is the first-order Marcum $Q$-function, and $G[\cdot|\cdot]$ represents the Meijer's $G-$function \cite[Eq. (9.301)]{tables}. Finally, $\mathcal{O}(\cdot)$ represents the Landau symbol; i.e., for two arbitrary functions $f(x)$ and $g(x)$, it holds that $f(x)=\mathcal{O}(g(x))$ when $|f(x)|\leq v |g(x)|\: \forall x\geq x_{0},\{v,x_{0}\}\in \mathbb{R}$.} 
\begin{align}
\nonumber
{\rm r}=\sqrt{\rm p}\bigg(&\sqrt{{\rm d}^{-\alpha_{\rm d}}_{\rm d}} h_{\rm d}+\sqrt{{\rm d}^{-\alpha_{1,1}}_{1,1}}\sqrt{{\rm d}^{-\alpha_{1,2}}_{1,2}}\mathbf{h}^{T}_{2}\mathbf{\Phi}_{1}\mathbf{h}_{1}\\
&+\sqrt{{\rm d}^{-\alpha_{2,1}}_{2,1}}\sqrt{{\rm d}^{-\alpha_{2,2}}_{2,2}}\mathbf{g}^{T}_{2}\mathbf{\Phi}_{2}\mathbf{g}_{1}\bigg)s+n,
\label{received}
\end{align}
where ${\rm p}$ is the transmit signal-to-noise ratio (SNR), $s\in \mathbb{C}$ is the unit-power transmit signal, $h_{\rm d}\in \mathbb{C}$ denotes the channel fading coefficient of the transceiver direct link; $\mathbf{h}_{1}\in \mathbb{C}^{N \times 1}$ and $\mathbf{h}_{2}\in \mathbb{C}^{N \times 1}$ are the channel vectors of the transmitter-to-${\rm RIS}_{1}$ and ${\rm RIS}_{1}$-to-receiver links, respectively; $\mathbf{\Phi}_{1}\triangleq \diag\{e^{{\rm j} \phi_{i}}\}^{N}_{i=1}$ denotes the phase rotations at ${\rm RIS}_{1}$; and $n\in \mathbb{C}$ defines the additive white Gaussian noise at the receiver such that $n\overset{\text{d}}=\mathcal{CN}(0,1)$. In practice, $\{\phi_{i}\}^{N}_{i=1}$ can only be configured from a given discrete-phase set $\mathcal{S}$, where $\mathcal{S}\in[-\pi,\pi]$; such that the cardinality of $\mathcal{S}$ is denoted by $|\mathcal{S}|=2^{q}$ with a $q-$bit quantization satisfying $q\geq 1$; yielding $\mathcal{S}_{l}\triangleq \{l \pi/2^{q-1}\}^{2^{q}-1}_{l=0}$. In the same analogy, $\mathbf{g}_{1}\in \mathbb{C}^{M \times 1}$, $\mathbf{g}_{2}\in \mathbb{C}^{M \times 1}$ and $\mathbf{\Phi}_{2}\in \mathbb{C}^{M \times M}$ are formulated, where $M$ denotes the number of passive elements at ${\rm RIS}_{2}$. Also, ${\rm d}_{\rm d}$, ${\rm d}_{1,1}$ and ${\rm d}_{1,2}$ denote the distances (in meters) between the direct link, transmitter-to-${\rm RIS}_{1}$ and ${\rm RIS}_{1}$-to-receiver links, respectively, with $\alpha_{\rm d}$, $\alpha_{1,1}$ and $\alpha_{1,2}$ their associated path-loss factors. Correspondingly, ${\rm d}_{2,1}$ and ${\rm d}_{2,2}$ denote the distances between the transmitter-to-${\rm RIS}_{2}$ and ${\rm RIS}_{2}$-to-receiver links, respectively, while $\alpha_{2,1}$ and $\alpha_{2,2}$ are their path-loss factors.

By applying coherent detection, i.e., given $\{h_{\rm d},\mathbf{h}_{1},\mathbf{h}_{2}\}$, the reference operator (say, operator-1) strives to align the received gains between the RIS-enabled and direct links so as to obtain the maximum achievable output. Then, according to \cite[Eq. (28)]{j:WuZhang2019}, the optimum setup of the $i^{\rm th}$ phase shift element of ${\rm RIS}_{1}$ becomes $\phi_{i}=e^{{\rm j} (\angle[h_{\rm d}]-\angle[\mathbf{h}_{1,i}]-\angle[\mathbf{h}_{2,i}])}$ where $i\in [1,N]$. Notably, operator-1 knows nothing about channels $\mathbf{g}_{1}$ and $\mathbf{g}_{2}$ since they are controlled by a different operator, which in turn cause the so-called IOI. Doing so, the post-detected signal at the receiver stems as  
\begin{align}
\nonumber
{\rm y}&=e^{-{\rm j} \angle[h_{\rm d}]}{\rm r}\\
\nonumber
&=\sqrt{\rm p}\bigg(\sqrt{{\rm d}^{-\alpha_{\rm d}}_{\rm d}} |h_{\rm d}|+\sqrt{{\rm d}^{-\alpha_{1,1}}_{1,1}}\sqrt{{\rm d}^{-\alpha_{1,2}}_{1,2}}\mathbf{h}^{T}_{2}\mathbf{\Theta}\mathbf{h}_{1}\\
&\ \ \ +\sqrt{{\rm d}^{-\alpha_{2,1}}_{2,1}}\sqrt{{\rm d}^{-\alpha_{2,2}}_{2,2}}e^{-{\rm j} \angle[h_{\rm d}]}\mathbf{g}^{T}_{2}\mathbf{\Phi}_{2}\mathbf{g}_{1}\bigg)s+e^{-{\rm j} \angle[h_{\rm d}]}n,
\label{detected}
\end{align}
where 
\begin{align}
\mathbf{\Theta}\triangleq \diag\left\{\exp\left({\rm j} \min_{l}\big[\left|\phi_{i}-\angle[h_{\rm d}]-\mathcal{S}_{l}\right|\big]\right)\right\}^{N}_{i=1}
\label{theta}
\end{align}
is the phase shift error matrix caused by the quantization mismatch between the ideal phase adjustment and practically feasible discrete-valued $\mathbf{\Phi}_{1}$. In essence, each entry of $\mathbf{\Theta}$ is zero-mean and can be modeled by a uniform distribution within the range $[-2^{-q}\pi,2^{-q}\pi]$. Also, $e^{-{\rm j} \angle[h_{\rm d}]}n\overset{\text{d}}=n$ due to the isotropic identity of circularly symmetric Gaussian RVs. Since coherent detection is considered at the receiver, as previously stated, operator-1 is aware of signals $\{h_{\rm d},\mathbf{h}_{1},\mathbf{h}_{2}\}$; i.e., a perfect channel-state information (CSI) is assumed. On the other hand, it does not control nor is aware of the CSI-related links regarding ${\rm RIS}_{2}$ (used by operator-2). The resultant received SNR at the reference user (subscribed to operator-1) yields as
\begin{align}
\nonumber
\gamma\triangleq \rm p \bigg|&\sqrt{{\rm d}^{-\alpha_{\rm d}}_{\rm d}} |h_{\rm d}|+\sqrt{{\rm d}^{-\alpha_{1,1}}_{1,1}}\sqrt{{\rm d}^{-\alpha_{1,2}}_{1,2}}\mathbf{h}^{T}_{2}\mathbf{\Theta}\mathbf{h}_{1}\\
&+\sqrt{{\rm d}^{-\alpha_{2,1}}_{2,1}}\sqrt{{\rm d}^{-\alpha_{2,2}}_{2,2}}e^{-{\rm j} \angle[h_{\rm d}]}\mathbf{g}^{T}_{2}\mathbf{\Phi}_{2}\mathbf{g}_{1}\bigg|^{2},
\label{snrrr}
\end{align}
where the two left-most sum terms of \eqref{snrrr} are considered as known channel gains, while the last term is the (unknown) source of randomness due to the presence of ${\rm RIS}_{2}$, which in turn causes a potentially strong fluctuation of the above SNR.

\section{Outage Probability and Spectral Efficiency}
Taking into consideration the fact that RISs are usually equipped with a vast number of low-cost passive elements, the well-known central limit theorem can be invoked which efficiently approximates the RIS-enabled channel entries of \eqref{detected} as Gaussian entries.\footnote{The Gaussianity assumption works well even with a moderately low RIS array $N=8$ \cite{j:YashvanthMurthy2023}.} Therefore, based on \cite[Lemma~1 and Corollary~2]{j:Badiu2020} and for ease of presentation, we introduce the following auxiliary variables
\begin{align}
\mathcal{X}\triangleq \sqrt{{\rm d}^{-\alpha_{1,1}}_{1,1}}\sqrt{{\rm d}^{-\alpha_{1,2}}_{1,2}}\mathbf{h}^{T}_{2}\mathbf{\Theta}\mathbf{h}_{1}\overset{\text{d}}\approx \mathcal{CN}\left(N \mu_{\mathcal{X}},N \mathcal{V}_{\mathcal{X}},N \mathcal{P}_{\mathcal{X}}\right),
\label{variableX}
\end{align}
and
\begin{align}
\mathcal{Y}\triangleq \sqrt{{\rm d}^{-\alpha_{2,1}}_{2,1}}\sqrt{{\rm d}^{-\alpha_{2,2}}_{2,2}}e^{-{\rm j} \angle[h_{\rm d}]}\mathbf{g}^{T}_{2}\mathbf{\Phi}_{2}\mathbf{g}_{1}\overset{\text{d}}\approx \mathcal{CN}\left(0,M \mathcal{V}_{\mathcal{Y}}\right),
\label{variableY}
\end{align}
where 
\begin{align}
\left\{\begin{array}{ll}\mu_{\mathcal{X}}=\sqrt{{\rm d}^{-\alpha_{1,1}}_{1,1}}\sqrt{{\rm d}^{-\alpha_{1,2}}_{1,2}} \vartheta_{1}A_{1}A_{2},\\
\mathcal{V}_{\mathcal{X}}={\rm d}^{-\alpha_{1,1}}_{1,1}{\rm d}^{-\alpha_{1,2}}_{1,2} \left(1-\vartheta^{2}_{1}A^{2}_{1}A^{2}_{2}\right),\\
\mathcal{V}_{\mathcal{Y}}={\rm d}^{-\alpha_{2,1}}_{2,1}{\rm d}^{-\alpha_{2,2}}_{2,2},\\
\mathcal{P}_{\mathcal{X}}={\rm d}^{-\alpha_{1,1}}_{1,1}{\rm d}^{-\alpha_{1,2}}_{1,2} \left(\vartheta_{2}-\vartheta^{2}_{1}A^{2}_{1}A^{2}_{2}\right),\end{array}\right.
\label{PX}
\end{align}
with $A_{l}\triangleq \mathbb{E}\left[|\mathbf{h}_{l}|\right]=\sqrt{\frac{\pi}{4 (\kappa_{l}+1)}}{}_1F_{1}\left(-\frac{1}{2},1;-\kappa_{l}\right),\quad l\in\{1,2\}$, $\kappa_{l}$ is the Rician $K-$factor of the $l^{\rm th}$ link and $\vartheta_{l}\triangleq \mathbb{E}\left[\mathbf{\Theta}^{l}\right]={\rm sinc}(2^{-q+l-1}\pi),\quad l\in\{1,2\}$. Finally, $\mathcal{Z}\triangleq \sqrt{{\rm d}^{-\alpha_{\rm d}}_{\rm d}} |h_{\rm d}|$ such that $f_{\mathcal{Z}}(x)=\frac{2 x}{\mathcal{V}_{\rm d}}e^{-x^{2}/\mathcal{V}_{\rm d}}$, where $\mathcal{V}_{\rm d}={\rm d}^{-\alpha_{\rm d}}_{\rm d}$.

Capitalizing on the above symbolism, we can compactly rewrite \eqref{detected} as ${\rm y}=\sqrt{\rm p}\Xi s+e^{-{\rm j} \angle[h_{\rm d}]}n$ with $\Xi \triangleq \mathcal{Z}+\mathcal{X}+\mathcal{Y}$, whereas the received SNR of the post-processed signal is captured by
\begin{align}
\gamma={\rm p}\left|\Xi\right|^{2}.
\label{snr}
\end{align}
Conditioned on the direct channel, the total channel fading is a non-circular complex-valued Gaussian RV distributed as
\begin{align}
\Xi\overset{\text{d}}=\mathcal{CN}\left(N \mu_{\mathcal{X}}+\mathcal{Z},N \mathcal{V}_{\mathcal{X}}+M \mathcal{V}_{\mathcal{Y}},N \mathcal{P}_{\mathcal{X}}+M \mathcal{V}_{\mathcal{Y}}\right),
\label{Xidistr}
\end{align}
having a real and imaginary part, respectively, modeled by \cite{j:Picinbono1996}
\begin{align}
{\rm Re}\{\Xi\}\overset{\text{d}}=&\mathcal{N}\bigg(N \mu_{\mathcal{X}}+\mathcal{Z},\underbrace{\frac{1}{2}\left(N \mathcal{P}_{\mathcal{X}}+N \mathcal{V}_{\mathcal{X}}+2 M \mathcal{V}_{\mathcal{Y}}\right)}_{\triangleq \sigma^{2}}\bigg),\\
\textrm{and }{\rm Im}\{\Xi\}\overset{\text{d}}=&\mathcal{N}\left(0,\frac{N}{2}\left(\mathcal{V}_{\mathcal{X}}-\mathcal{P}_{\mathcal{X}}\right)\right).
\label{real}
\end{align}
We proceed by analyzing the cumulant generating function (CGF) of $|\Xi|^{2}={\rm Re}\{\Xi\}^{2}+{\rm Im}\{\Xi\}^{2}$, conditioned on $\mathcal{Z}$, and further approximating it according to \cite[Eq. (20)]{j:Badiu2020} as 
\begin{align}
\nonumber
&\mathcal{K}_{|\Xi|^{2}|\mathcal{Z}}(t)={\rm ln}\mathbb{E}[e^{t |\Xi|^{2}}]={\rm ln}\mathbb{E}[e^{t {\rm Re}\{\Xi\}^{2}}]+{\rm ln}\mathbb{E}[e^{t {\rm Im}\{\Xi\}^{2}}]\\
\nonumber
&=\frac{t (N \mu_{\mathcal{X}}+\mathcal{Z})^{2}}{1-2 \sigma^{2} t}-\frac{1}{2}{\rm ln}(1-2 \sigma^{2} t)\\
\nonumber
&\ \ \ -\frac{1}{2}{\rm ln}(1-N\left(\mathcal{V}_{\mathcal{X}}-\mathcal{P}_{\mathcal{X}}\right) t)\\
&=-\frac{t (N \mu_{\mathcal{X}}+\mathcal{Z})^{2}}{4 \sigma^{2}}{\rm ln}(1-4 \sigma^{2} t)+\mathcal{O}(N^{-1}),
\end{align}
which is the CGF of a (weighted) Gamma distributed RV with a shape $(N \mu_{\mathcal{X}}+\mathcal{Z})^{2}/(4 \sigma^{2})$ and scale $4 \sigma^{2}$. The corresponding unconditional CGF arises as 
\begin{align}
\nonumber
\mathcal{K}_{|\Xi|^{2}}(t)&\approx-\frac{t \int^{\infty}_{0}(N \mu_{\mathcal{X}}+y)^{2}f_{\mathcal{Z}}(y){\rm d}y}{4 \sigma^{2}}{\rm ln}(1-4 \sigma^{2} t)\\
&=-\frac{t \left(N^{2} \mu^{2}_{\mathcal{X}}+\mathcal{V}_{\rm d}+\sqrt{\pi \mathcal{V}_{\rm d}}N \mu_{\mathcal{X}}\right)}{4 \sigma^{2}}{\rm ln}(1-4 \sigma^{2} t).
\end{align}

Putting altogether the latter results and based on \eqref{snr}, the CDF of the received SNR is derived after some straightforward manipulations as
\begin{align}
F_{\gamma}(x)\approx 1-\frac{\Gamma\left(m_{N},\frac{x}{{\rm p}\overline{\gamma}}\right)}{\Gamma\left(m_{N}\right)},
\label{cdfsnr}
\end{align}
where $m_{N}\triangleq (N^{2} \mu^{2}_{\mathcal{X}}+\mathcal{V}_{\rm d}+\sqrt{\pi \mathcal{V}_{\rm d}}N \mu_{\mathcal{X}})/\overline{\gamma}$, $\overline{\gamma}\triangleq 4 \sigma^{2}$ and $\sigma^{2}$ is defined back in \eqref{real}. Thus, outage probability at a given SNR threshold, $P_{\rm out}(\gamma_{\rm th})$, can be directly computed since $P_{\rm out}(\gamma_{\rm th})=F_{\gamma}(\gamma_{\rm th})$. Using the property $\Gamma(a,z)\rightarrow \Gamma(a)-z^{a}/a$, as $z\rightarrow 0^{+}$, and assuming a bounded transmit power and as the RIS arrays grow extremely high (i.e., $\{N,M\}\rightarrow \infty$), it is not difficult to show that
\begin{align}
P^{(\{N,M\}\rightarrow \infty)}_{\rm out}(\gamma_{\rm th})\rightarrow \frac{1}{\Gamma(m_{N}+1)}\left(\frac{\gamma_{\rm th}}{{\rm p}\overline{\gamma}}\right)^{m_{N}}\propto \left({\rm p}\overline{\gamma}\right)^{-\frac{N^{2}}{M}},
\label{outageasympt}
\end{align}
thereby yielding a diversity order of $N^{2}/M$.

The spectral efficiency (in bps/Hz) is computed as
\begin{align}
\nonumber
\mathcal{C}&\triangleq \mathbb{E}[{\rm log}_{2}(1+\gamma)]=\frac{1}{{\rm ln}(2)}\int^{\infty}_{0}{\rm ln}(1+x)f_{\gamma}(x){\rm d}x\\
\nonumber
&=\frac{\left({\rm p}\overline{\gamma}\right)^{-m_{N}}}{{\rm ln}(2)\Gamma(m_{N})}\int^{\infty}_{0}{\rm ln}(1+x)x^{m_{N}-1}\exp\left(-\frac{x}{{\rm p}\overline{\gamma}}\right){\rm d}x\\
&=\frac{1}{{\rm ln}(2)\Gamma(m_{N})}G^{1,3}_{3,2}\left[{\rm p}\overline{\gamma}~\vline
\begin{array}{c}
1-m_{N},1,1 \\
1,0
\end{array}\right],
\label{capacity}
\end{align}
where the last equality arises by directly utilizing \cite[Eqs. (8.4.6.5) and (2.24.3.1)]{b:prudnikovvol3}. Although \eqref{capacity} is straightforward, Meijer's $G-$function is included which is cumbersome; to this end, we further resort to an asymptotic approximation to gain more insights. Notice that $\overline{\gamma}\rightarrow \infty$ as $N$ grows unboundedly. Thereby, taking the Taylor expansion of \eqref{capacity} and keeping only the zeroth (most dominant) term, we get quite a simple approximation expressed as
\begin{align}
\mathcal{C}_{(N\rightarrow \infty)}\rightarrow {\rm ln}({\rm p}\overline{\gamma}) \psi({\rm p}\overline{\gamma})/{\rm ln}(2). 
\label{asycap}
\end{align}

\subsection{Special Cases}
\subsubsection{Zero phase shift error}
This scenario corresponds when $\mathbf{\Theta}=\mathbf{0}$, which can be realized when a perfect phase shift alignment occurs and/or $q\rightarrow \infty$. In this case, \eqref{cdfsnr} and \eqref{capacity} also hold by accordingly substituting $\vartheta_{1}=\vartheta_{2}=1$ while setting $\sigma^{2}=N \mathcal{V}_{\mathcal{X}}+M\mathcal{V}_{\mathcal{Y}}$.

\subsubsection{w/o direct link}
When the direct link is negligible or absent, $h_{\rm d}=0$; thus all the aforementioned results hold by simply setting $\mathcal{V}_{\rm d}=0$. Particularly, this affects $m_{N}$ within \eqref{cdfsnr}, \eqref{outageasympt} and \eqref{capacity} which becomes $m_{N}=N^{2} \mu^{2}_{\mathcal{X}}/\overline{\gamma}$. 

\subsubsection{w/o RIS at reference operator}
For the special case where $N=0$, which describes the scenario when operator-1 does not use RIS while operator-2 does, the total channel fading coefficient as per \eqref{Xidistr} reduces to $\Xi\overset{\text{d}}=\mathcal{N}\left(\mathcal{Z},M \mathcal{V}_{\mathcal{Y}}\right)$. Thus, the total channel fading coefficient is turned to a circular-symmetric Gaussian RV, which requires a different analytical approach. In fact, the received SNR, conditioned on the direct channel, is a non-central chi-squared RV with one degree-of-freedom (DoF). Its CDF is given by $F^{(N=0)}_{\gamma|\mathcal{Z}}(x)=1-Q_{1}(\sqrt{\frac{2 \mathcal{Z}^{2}}{M \mathcal{V}_{\mathcal{Y}}}},\sqrt{\frac{2 x}{{\rm p} M \mathcal{V}_{\mathcal{Y}}}})$. The corresponding unconditional CDF is obtained as
\begin{align}
\nonumber
&F^{(N=0)}_{\gamma}(x)=1-\int^{\infty}_{0}Q_{1}\left(\sqrt{\frac{2 y^{2}}{M \mathcal{V}_{\mathcal{Y}}}},\sqrt{\frac{2 x}{{\rm p} M \mathcal{V}_{\mathcal{Y}}}}\right)f_{\mathcal{Z}}(y){\rm d}y\\
\nonumber
&=1-\exp\left(-\frac{x}{{\rm p} M \mathcal{V}_{\mathcal{Y}}}\right)-\exp\left(\scriptstyle-\frac{x}{{\rm p} M \mathcal{V}_{\mathcal{Y}}\mathcal{V}_{\rm d}\left(\frac{1}{M \mathcal{V}_{\mathcal{Y}}}+\frac{1}{\mathcal{V}_{\rm d}}\right)}\displaystyle\right)\\
&\ \ \ \times \left[1-\exp\left(-\frac{x}{{\rm p} (M \mathcal{V}_{\mathcal{Y}})^{2}\left(\frac{1}{M \mathcal{V}_{\mathcal{Y}}}+\frac{1}{\mathcal{V}_{\rm d}}\right)}\right)\right],
\label{snrn0cdf}
\end{align}
where the last equality relies on \cite[Lemma~1]{j:sofotasiosmarcum}. Obviously, in the absence of RIS-enabled environment from every operator (i.e., when both $\{N,M\}=0$), \eqref{snrn0cdf} reduces to the trivial case of Rayleigh-only channel fading caused by the direct link yielding $F_{\gamma}(x)=1-\exp(-x/({\rm p}\mathcal{V}_{\rm d}))$. On the other hand, when $M\rightarrow \infty$, \eqref{snrn0cdf} tends to $F^{(M\rightarrow \infty)}_{\gamma}(x)\rightarrow x/({\rm p} M \mathcal{V}_{\mathcal{Y}})\rightarrow 0^{+}$. The latter results reveal that the presence of RIS from other operators (\emph{yet}, when the reference operator does not implement RIS for signal propagation) is beneficial for the system performance when $M$ is large; which is in agreement with \cite{c:YashvanthMurthy2023,j:yashvanth2023}.

The spectral efficiency is expressed as
\begin{align}
\nonumber
\mathcal{C}_{(N=0)}&\triangleq \mathbb{E}[{\rm log}_{2}(1+\gamma)]=\frac{1}{{\rm ln}(2)}\int^{\infty}_{0}\frac{1-F_{\gamma}(x)}{1+x}{\rm d}x\\
\nonumber
&=\frac{1}{{\rm ln}(2)}\Bigg\{\exp\scriptstyle\left(\frac{1}{{\rm p} M \mathcal{V}_{\mathcal{Y}}+{\rm p} \mathcal{V}_{\rm d} \mathcal{V}^{2}_{\mathcal{Y}}}\displaystyle\right)\Bigg[\Gamma\scriptstyle\left(0,\frac{1}{{\rm p} M \mathcal{V}_{\mathcal{Y}}+{\rm p} \mathcal{V}_{\rm d} \mathcal{V}^{2}_{\mathcal{Y}}}\displaystyle\right)\\
\nonumber
&\ \ \ -\exp\scriptstyle\left(\frac{\mathcal{V}_{\rm d}}{{\rm p} (M \mathcal{V}_{\mathcal{Y}})^{2}+{\rm p} \mathcal{V}_{\rm d} \mathcal{V}^{3}_{\mathcal{Y}}}\right)\displaystyle\Gamma\scriptstyle\left(0,\frac{\mathcal{V}_{\rm d}+M \mathcal{V}_{\mathcal{Y}}}{{\rm p} M \mathcal{V}^{2}_{\mathcal{Y}}(M+\mathcal{V}_{\rm d} \mathcal{V}_{\mathcal{Y}})}\right)\Bigg]\\
&\ \ \ +\exp\scriptstyle\left(\frac{1}{{\rm p} M \mathcal{V}_{\mathcal{Y}}}\displaystyle\right)\Gamma\scriptstyle\left(0,\frac{1}{{\rm p} M \mathcal{V}_{\mathcal{Y}}}\displaystyle\right)\Bigg\},
\label{capacity2}
\end{align}
where the last equation arises by utilizing \cite[Eq. (3.383.10)]{tables}.

\subsubsection{Multi-operators/multi-RISs}
This scenario corresponds to the case when there are more than two heterogeneous RISs. Specifically, consider multiple operators, where each controls its own RIS, operate in a close vicinity at a target area. Then, following similar lines of reasoning as for the derivation of \eqref{variableY} and using the linear property of Gaussian RVs, the channel fading caused by the multi-operator RIS-enabled links reads as 
\begin{align}
\nonumber
\mathcal{Y}&\triangleq \sum^{U}_{u=1}\sqrt{{\rm d}^{-\alpha_{u,1}}_{u,1}}\sqrt{{\rm d}^{-\alpha_{u,2}}_{u,2}}e^{-{\rm j} \angle[h_{\rm d}]}\mathbf{g}^{T}_{2,u}\mathbf{\Phi}_{2,u}\mathbf{g}_{1,u}\\
&\overset{\text{d}}\approx \mathcal{CN}\left(0,\sum^{U}_{u=1}M_{u} \mathcal{V}_{\mathcal{Y},u}\right),
\label{variableYmulti}
\end{align}
where $U$ represents the number of inter-operator RISs, the $u^{\rm th}$ RIS is equipped with $M_{u}$ RIS elements as well as $\mathcal{V}_{\mathcal{Y},u}$ corresponds to the $u^{\rm th}$ RIS and is given in \eqref{PX}. It is not difficult to show that the previous analytical results are directly extended to this general case, by simply substituting $M \mathcal{V}_{\mathcal{Y}}$ with $\sum^{U}_{u=1}M_{u} \mathcal{V}_{\mathcal{Y},u}$ in $\sigma^{2}$ at \eqref{cdfsnr}, \eqref{outageasympt} and \eqref{capacity}.

\subsection{Takeaways}
Apart from the above analytical results, a set of useful outcomes, and more so, some key engineering insights can be extracted: (a.) At the presence of IOI, the channel (diversity) gain provided by RIS-enabled communication is no longer quadratic to the number of RIS elements. In fact, it is reduced by a factor depending on the number of RIS elements of other operators' RISs. For an identical channel environment of all operators around the reference user, such a gain is $\propto N^{2}/\sum^{U}_{u=1}M_{u}$. This effect can be interpreted as an incoherent fluctuation of the coherent detection at the receiver; (b.) From a fair scheduling standpoint, a suitable regulation dictates that \emph{all operators should have equal number of RIS elements} at their RIS deployments (i.e., $N=M_{u}\forall u$). In doing so, all the subscribers will experience a \emph{linear} gain $\propto N$ and a `\emph{flat}' (say, identical) IOI. 

\section{Numerical Results and Discussion}
In this section, the derived analytical results are verified via numerical validation (in line-curves), whereas they are cross-compared with corresponding Monte-Carlo simulations (in solid-circle marks). Without loss of generality and for the sake of clarity, we set the quantization bit size $q=3$, the path losses $\{{\rm d}^{-\alpha_{\rm d}}_{\rm d},{\rm d}^{-\alpha_{1,1}}_{1,1},{\rm d}^{-\alpha_{1,2}}_{1,2}\}=\{100^{-3.1},30^{-2.2},30^{-2.4}\}$ with respect to the links of reference operator (that is, operator-1) as well as $\{{\rm d}^{-\alpha_{2,1}}_{2,1},{\rm d}^{-\alpha_{2,2}}_{2,2}\}=\{30^{-2.2},30^{-2.4}\}$ regarding operator-2; which are typical for a dense urban terrestrial with LoS or near LoS signal propagation. The Rician $K-$factors of the transmitter-to-${\rm RIS}_{1}$ and ${\rm RIS}_{1}$-to-receiver links are defined, respectively, as $\kappa_{1}=10$ and $\kappa_{2}=6$.

\begin{figure}[!t]
\centering
\includegraphics[trim=1.5cm .0cm 1.5cm 0.0cm, clip=true,totalheight=0.4\textheight]{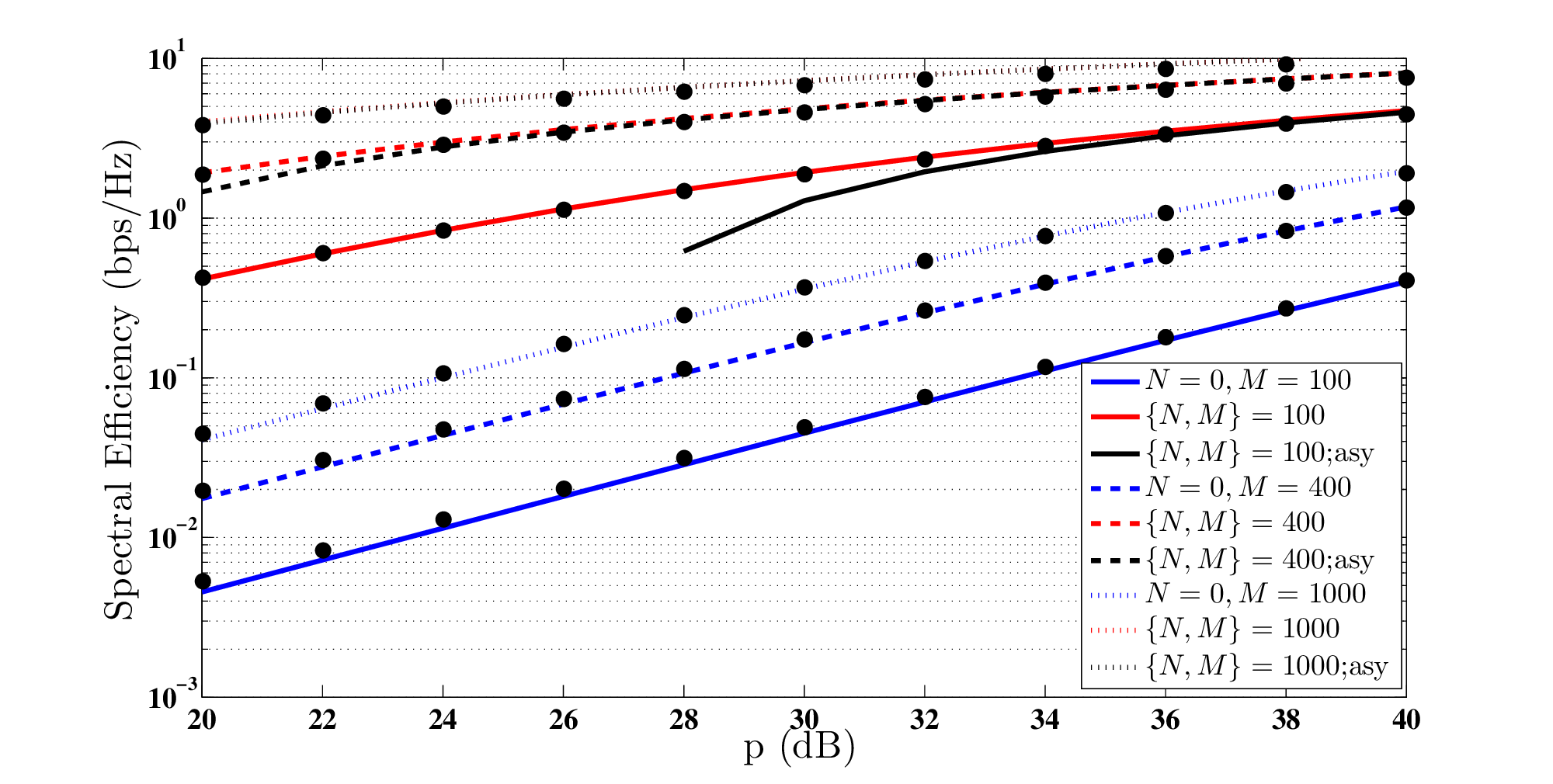}
\caption{Spectral efficiency vs. various transmit SNR values for different system setups. The term `${\rm asy}$' stands for the asymptotic $\mathcal{C}_{(N\rightarrow \infty)}$ as per \eqref{asycap}.}
\label{fig2}
\end{figure}

In Fig.~\ref{fig2}, two different system scenarios are considered; \emph{Case-1} where a RIS configuration is established only at the external operator (with a scaled range of RIS array $M$) and \emph{Case-2} where both operators (reference and external) are equipped with equal-sized yet scaled-range RISs. Obviously, the presence of an external RIS at Case-1 (illustrated in blue curves) is beneficial for the system performance and this gets more emphatic when $M$ increases. Although incoherently, the presence of an (uncontrolled) RIS provides a rich-scattered surrounding and hence introduces new DoF for the received signal. Nevertheless, as expected, when the reference operator is aided with its own RIS (Case-2 in red curves), the system performance is considerably being enhanced due to the coherent channel gain detection employed.  

\begin{figure}[!t]
\centering
\includegraphics[trim=1.5cm .0cm 1.5cm 0.0cm, clip=true,totalheight=0.4\textheight]{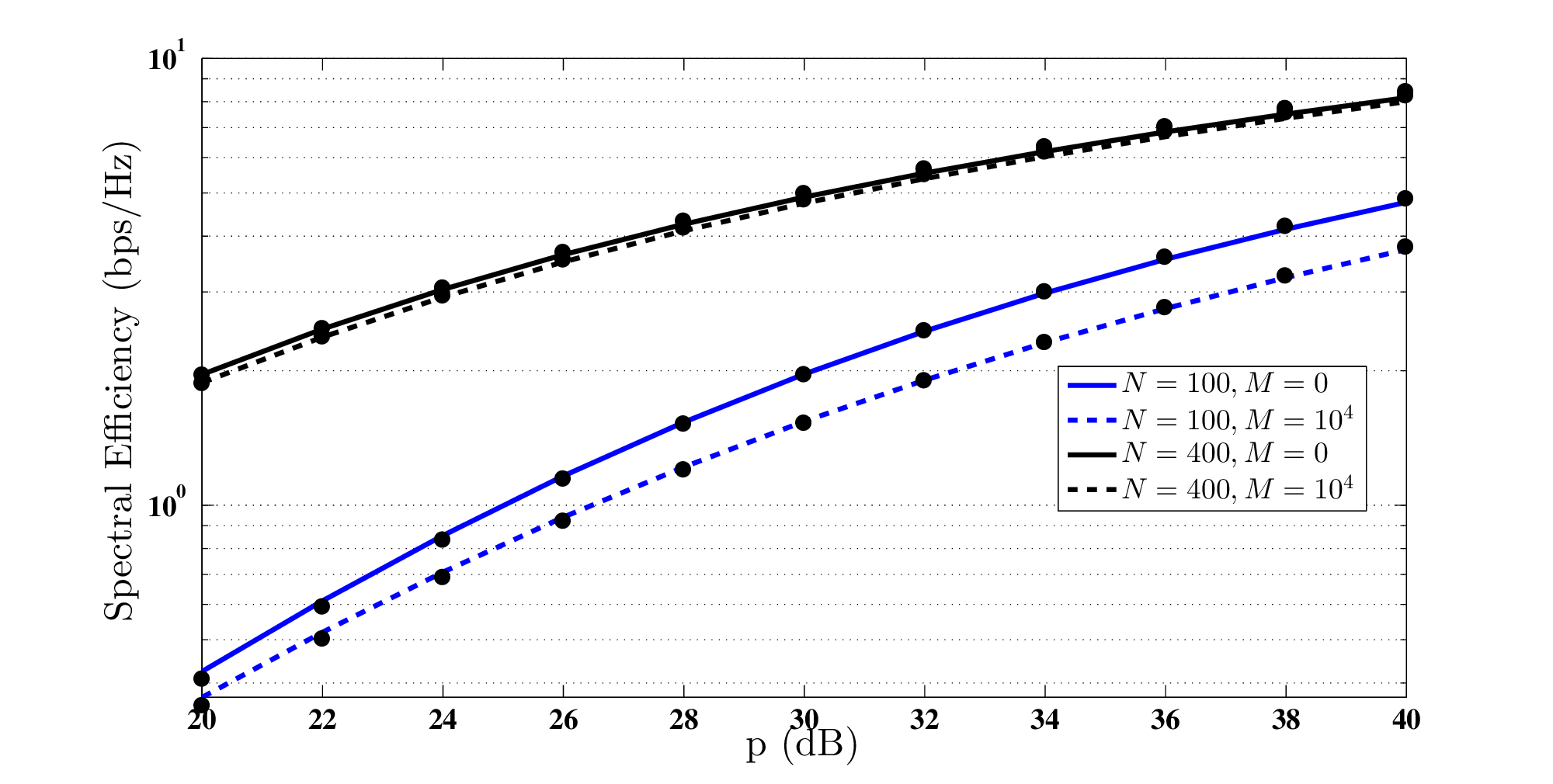}
\caption{Spectral efficiency vs. various transmit SNR values of the considered system model while using different ranges of the heterogeneous RIS array.}
\label{fig3}
\end{figure}

In Fig.~\ref{fig3}, the system performance is further evaluated from another perspective. Specifically, we set the number of RIS elements for the reference operator at a fixed value; namely $N=100$ or $N=400$ (e.g., a rectangular $10\times 10$ or $20\times 20$ array, correspondingly) over different ranges of the external RIS. To gain impactful insights, two extreme cases are illustrated; $M=0$ which corresponds to the absence of IOI and $M=10^{4}$ (e.g., a vast $100\times 100$ array) which reflects on an enormously larger RIS with respect to the reference one. Apparently, the presence of RIS-based IOI deteriorates the system performance; however it is worthy to state that the performance difference is marginal as $N$ gets higher. Thereby, if a certain operator is equipped with a large RIS, its communication performance is slightly influenced by a nearby heterogeneous RIS, even when the latter (external) RIS array is larger that the reference one.     

\section{Conclusion}
The effect of IOI in a multi-operator / multi-RIS communication system was analytically studied. The proposed approach holds for an arbitrary range of channel fading conditions, while includes the (potential) impact of the direct link channel gain and practical situation of discrete phase shifts at RIS. The revealed outcomes showcase that IOI may slightly influence the system performance under the conditions specified into this paper. From a fairness viewpoint, an effective solution is to deploy equal-sized RIS arrays for all the active network providers.

%\balance

\bibliographystyle{IEEEtran}
\bibliography{IEEEabrv,References}

\vfill

\end{document}